\documentclass[a4paper,fleqn]{cas-sc}

\usepackage[numbers]{natbib}
\usepackage{subcaption}

\graphicspath{{figs/}}

\def\tsc#1{\csdef{#1}{\textsc{\lowercase{#1}}\xspace}}
\tsc{WGM}
\tsc{QE}

\begin{document}
\let\WriteBookmarks\relax
\def\floatpagepagefraction{1}
\def\textpagefraction{.001}

\shorttitle{}    

\shortauthors{L. Nulandaya et~al.}  

\title [mode = title]{Large transformation strain of Ni$_{50}$Fe$_{23}$Ga$_{27}$ microwires produced by the Taylor-Ulitovsky method}  

\author[1]{Limpat Nulandaya}[orcid=0000-0002-0898-7933]
\cormark[1]
\ead{nulandaya@fzu.cz}

\author[1]{Petr Cejpek}[orcid=0000-0001-7745-325X]
\ead{cejpekp@fzu.cz}

\author[1]{Ladislav Straka}[orcid=0000-0003-1136-4996]
\ead{ladislav.straka@fzu.cz}

\author[1]{Oleg Heczko}[orcid=0000-0002-7647-3141]
\ead{heczko@fzu.cz}

\author[2]{Rastislav Varga}[orcid=0000-0003-2153-1065]
\ead{varga@rvmagnetics.com}

\affiliation[1]{
    organization={Institute of Physics of the Czech Academy of Sciences},
    city={Prague},
    postcode={18200},
    country={Czech Republic}
}

\affiliation[2]{
    organization={RVactuators, s.r.o.},
    city={Dolni Brezany},
    postcode={25241},
    country={Czech Republic}
}

\cortext[1]{Corresponding author}



\begin{abstract}
The martensitic transformation and associated tensile strain of Ni$_{50}$Fe$_{23}$Ga$_{27}$ microwires produced by the Taylor--Ulitovsky method were investigated using thermomechanical analysis, electron backscatter diffraction (EBSD), and in-situ tensile X-ray diffraction. EBSD mapping at room temperature reveals cubic austenite in a nonstrained part of microwire and tetragonal martensite in the strained part. Both EBSD and XRD show that both austenite and martensite have [001] directions approximately parallel to the wire axis. The axial transformation strain calculated from the measured lattice parameters is approximately $15.4\,\%$. Thermal cycling under constant stresses of 20--100 MPa reveals experimental transformation strains of 13.2--14.0$\,\%$. Increasing tensile stress raises the transformation temperatures, with a stress–temperature slope of approximately 1.52~MPa$/^{\circ}$C, comparable to that reported for [001]-oriented Ni--Fe--Ga single crystals. Overall the observation confirm that the large macroscopic strain of up to $14.0\,\%$ relates to the nearly complete martensite transformation in the microwire. \nocite{*}
\end{abstract}

\begin{graphicalabstract}
\includegraphics[width=0.85\linewidth]{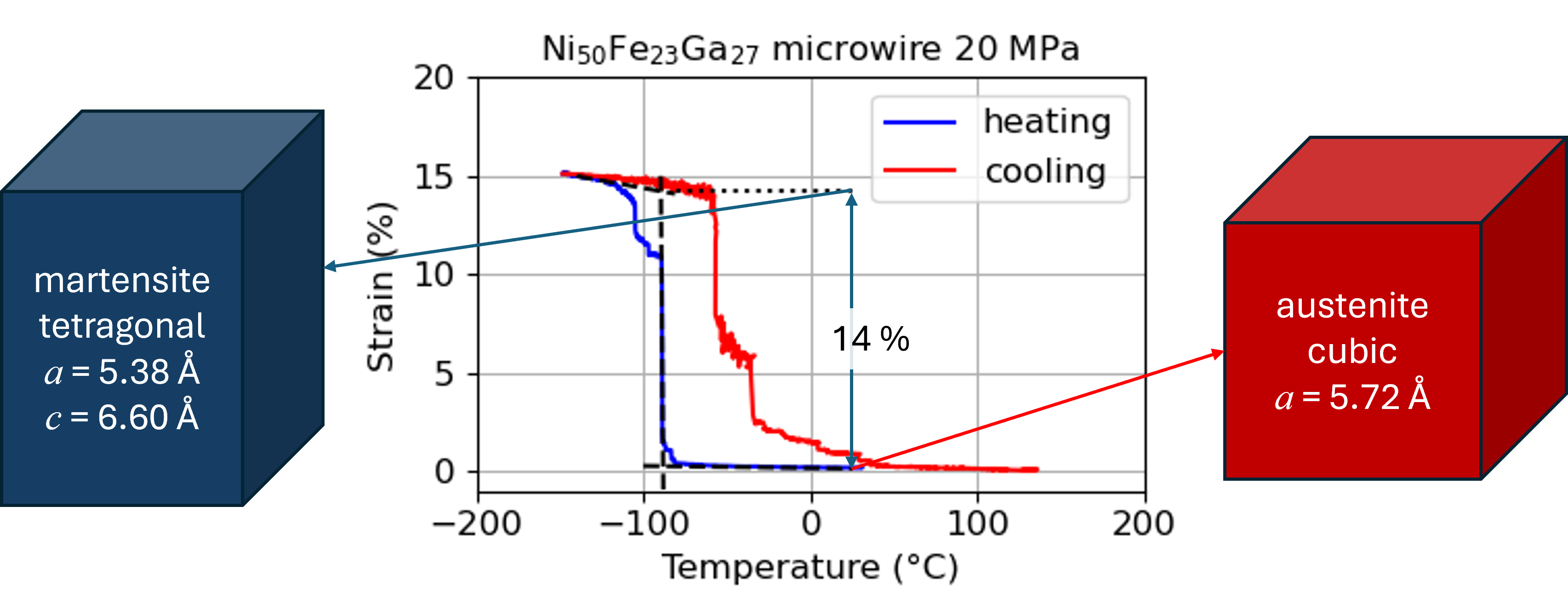}
\end{graphicalabstract}

\begin{highlights}
\item Ni$_{50}$Fe$_{23}$Ga$_{27}$ microwires show an experimental tensile transformation strain of about $14.0\,\%$.
\item Austenite and martensite have preferred crystallographic orientation [001] axis parallel to the wire’s axis.
\item The lattice parameters of austenite and martensite predict an~axial transformation strain of about $15.4\,\%$.
\end{highlights}


\maketitle

\section{Introduction}

Taylor-Ulitovsky glass-coated microwires represent a unique class of metallic materials produced through the simultaneous drawing of a molten metallic core enclosed within a softened glass capillary \cite{Taylor1924,Zhao2013}. This technique enables the fabrication of continuous composite microwires with metallic core diameters ranging from submicrometer dimensions to several tens of micrometers while maintaining excellent geometric control and reproducibility \cite{Baranov2017}. Microwires from the Taylor-Ulitovsky method exhibit microstructures that can differ significantly from those obtained by conventional bulk casting routes \cite{Chen2020,Chen2022,Salaheldeen2024}. The method allows continuous production of wire lengths extending to several kilometers, making it attractive for large-scale manufacturing \cite{Baranov2017}. These advantages have stimulated increasing interest in Taylor-Ulitovsky microwires for sensing, actuation, magnetic, and shape-memory applications, where miniaturization and high functional performance are essential \cite{Baranov2017,Chen2011,Frolova2021,Hennel2022}.

Among the materials processed using the Taylor-Ulitovsky technique, Ni--Fe--Ga ferromagnetic shape memory alloys (FSMAs) have attracted considerable attention because they combine thermoelastic martensitic transformations with magnetic-field-responsive properties \cite{Hamilton2007,Hennel2020,Frolova2021}. The martensitic transformation behavior of this alloys is highly sensitive to the chemical composition, crystallographic orientation, grain structure, and phase constitution \cite{Efstathiou2007,Heczko2008,Chen2011,Ueland2012,Hennel2020,Zhao2024}. Reversible transformations between the ordered cubic L2$_1$ austenitic phase and modulated martensitic structures provide the basis for their shape-memory behavior \cite{Hamilton2007,Frolova2021}, while their magnetic coupling enables additional functionalities such as magnetocaloric effects \cite{Yu2009,Hennel2022}. 

The importance of crystallographic orientation for transformation behavior has been demonstrated in single-crystalline Ni--Fe--Ga alloys. For example, Ni$_{54}$Fe$_{19}$Ga$_{27}$ single crystals oriented along the [001] direction were reported to exhibit exceptional cyclic stability, surviving more than 13,000 loading cycles at a tensile strain range of 10\,\%~\cite{Efstathiou2007}. Such results highlight the strong relationship between crystallographic texture and functional reliability in ferromagnetic shape-memory alloys. Pataky et al. reported Clausius--Clapeyron slopes of 4.05~MPa$/^{\circ}$C and 1.42~MPa$/^{\circ}$C for tensile loading along [011] and [001], respectively, in Ni$_{54}$Fe$_{19}$Ga$_{27}$ single crystals~\cite{Pataky2015}. In the case of Ni--Fe--Ga microwire, it was reported that the Ni$_{50}$Fe$_{25}$Ga$_{25}$ transformation strain $2\,\%$ with the transformation from cubic austenite ([111] axis parallel to the wire’s axis) to monoclinic martensite ([017] axis parallel to the wire’s axis).

Among the investigated compositions, Ni$_{50}$Fe$_{23}$Ga$_{27}$ is of particular interest because it is close to the stoichiometric composition of the ordered intermetallic Ni$_2$FeGa phase~\cite{Nath2015}. This composition was also reported to have unusually high magnetic anisotropy of austenite in Ni$_{50}$Fe$_{23}$Ga$_{27}$ ribbons, exceding that of martensite~\cite{Qian2011}. Despite these attractive characteristics, fundamental information regarding the crystallographic evolution and thermomechanical behavior for this composition under tensile loading remains limited.

In the present work, the thermomechanical response and crystallographic evolution of Ni$_{50}$Fe$_{23}$Ga$_{27}$ microwires subjected to tensile loading are investigated using electron backscatter diffraction (EBSD), in-situ tensile XRD, and complementary thermomechanical characterization by DMA. Particular emphasis is placed on comparing the axial crystallographic alignment of austenite and stress-induced martensite and relating the measured deformation to the lattice distortion.

\section{Experimental}
Ni-Fe-Ga microwires were drawn by the Taylor-Ulitovsky method~\cite{Baranov2017,Alam2020} and collected on a rolling bobbin. The master alloy for microwire production was prepared from high purity elements Ni, Fe, and Ga ($99.9\,\%$ Alfa Aesar$\textsuperscript{\texttrademark}$) by arc melter under an argon atmosphere. Simax$\textsuperscript{\textregistered}$ glass tube was used to obtain a glass-coated microwire during the production. The microwire selected for characterization was taken from a section in which the glass coating could be readily removed by pulling. Hereafter, “microwire” refers to the metallic core after removal of the glass coating.

\begin{figure}[h]
	\centering
	\begin{subfigure}[b]{0.48\linewidth}
		\centering
		\includegraphics[width=\linewidth]{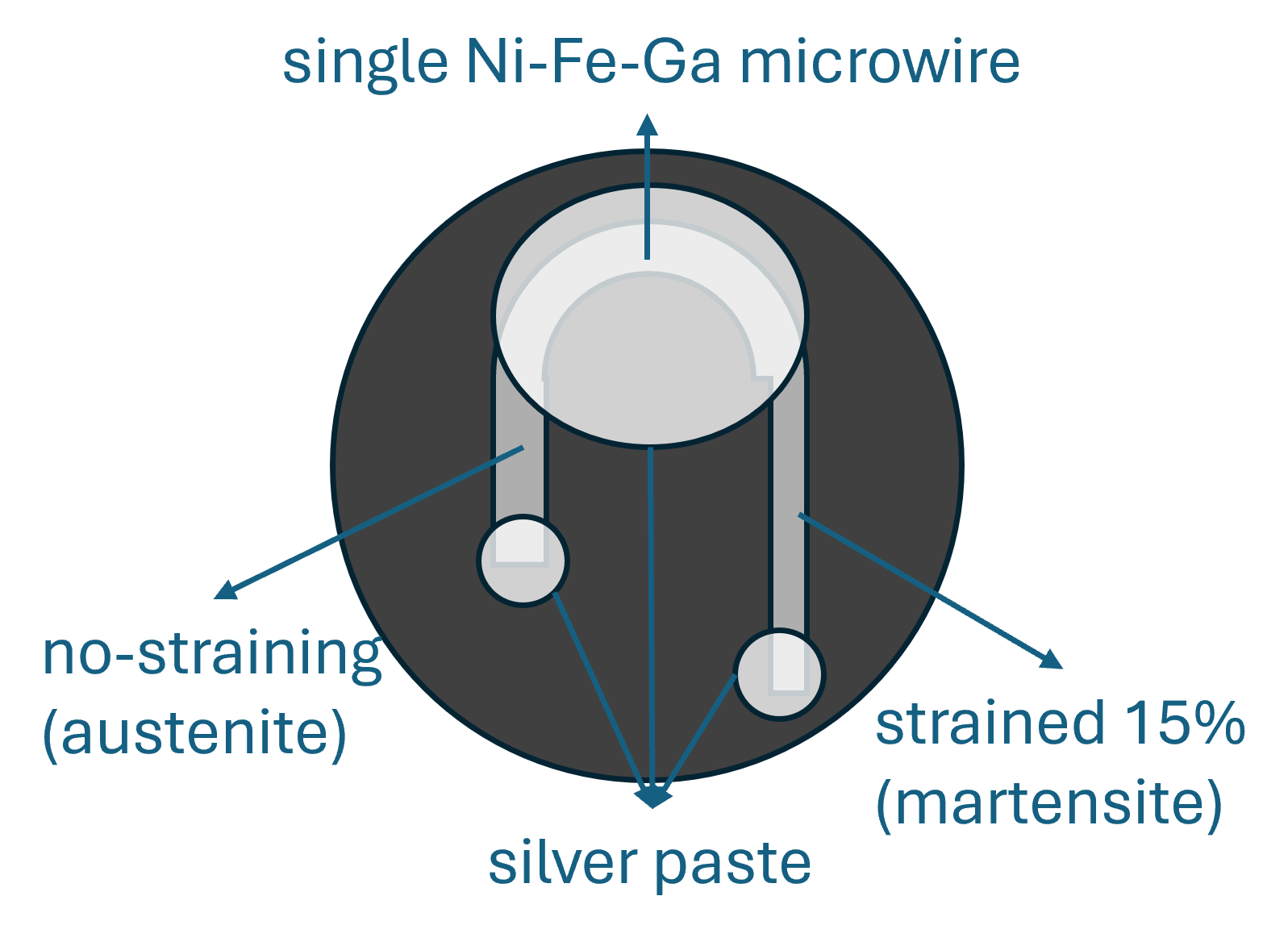}
		\caption{}
		\label{fig:picE1_a}
	\end{subfigure}
	\hfill
	\begin{subfigure}[b]{0.48\linewidth}
		\centering
		\includegraphics[width=\linewidth]{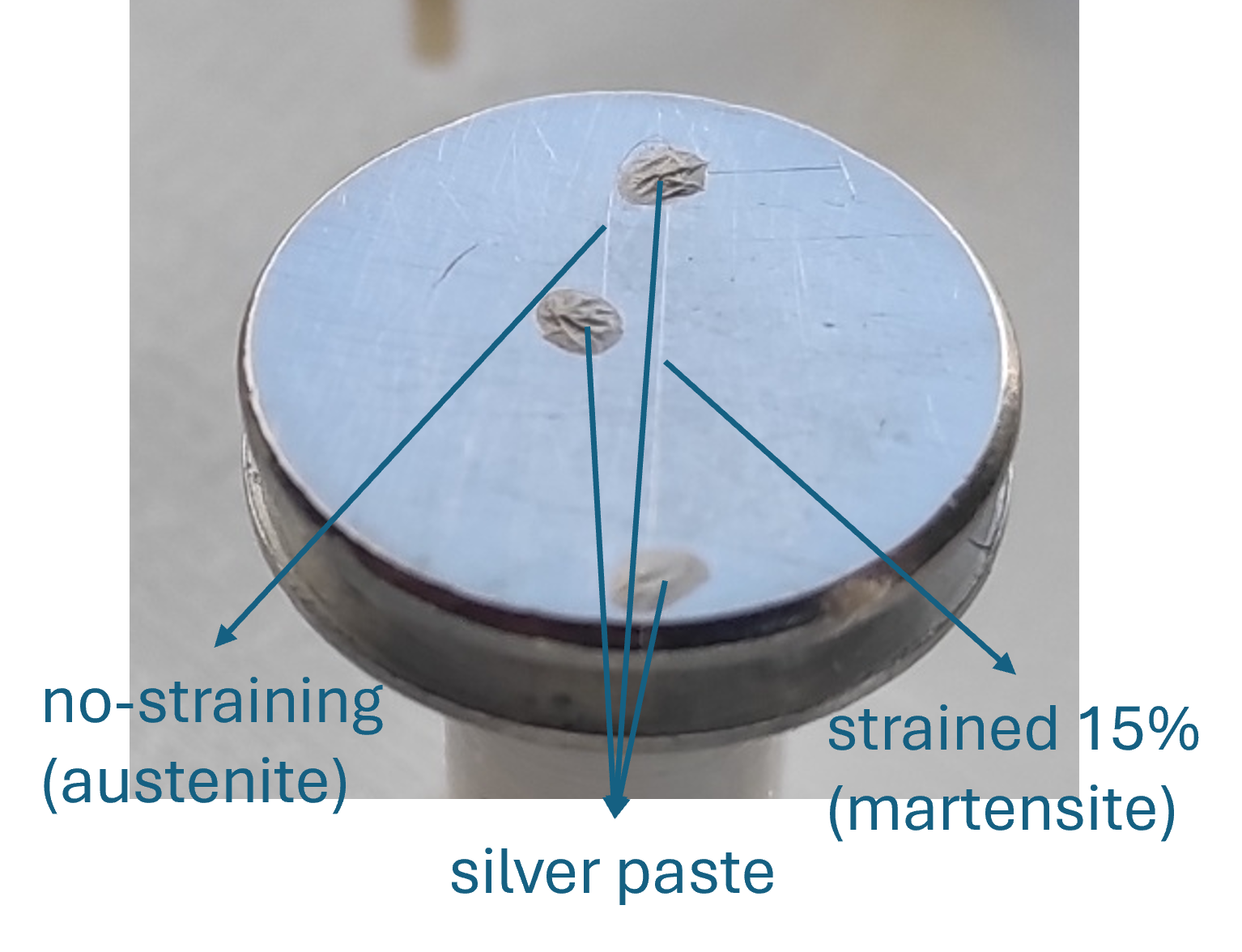}
		\caption{}
		\label{fig:picE1_b}
	\end{subfigure}
	\caption{(a) Illustration and (b) photo of a single microwire mounted on an SEM stub to obtain martensitic (by $15\,\%$ tensile strain) and austenitic regions}
	\label{fig:picE1}
\end{figure}

A~single Ni$_{50}$Fe$_{23}$Ga$_{27}$ microwire was mounted on an SEM stub to enable direct comparison between the austenitic and strain-induced martensitic states within the same specimen as in Fig.~\ref{fig:picE1}. The martensitic region was prepared by applying a~tensile strain of approximately $15\,\%$ to part of the initially austenitic microwire, followed by a fixation using conductive silver paste to preserve the deformation state during subsequent characterization. The selected strain level was determined from prior transformation strain, which indicated that $15\,\%$ strain was sufficient to fully induce the martensitic phase in the investigated microwire. The mounted microwire on SEM stub was examined using a 70$^\circ$ pretilted holder. Electron backscatter diffraction (EBSD, Oxford Instruments Symmetry S3) and energy-dispersive X-ray spectroscopy (EDX, Oxford Instruments Ultim Max Infinity) analyses were performed using a Thermo Fisher Scientific Apreo 2S LoVac Scanning Electron Microscope (SEM). All measurements were conducted at an accelerating voltage of 20~kV. Data acquisition was performed using AZtec 6.2. EBSD patterns were recorded with a resolution of $156 (width) \times 128 (height)$ pixels. After acquisition, the complete EBSD and EDS datasets were exported in Oxford Instruments H5OINA format for subsequent processing. Phase orientation mapping for martensite and austenite parts in microwire were processed using AZtecCrystal 3.3. For the EBSD patterns fitting of the austenite and stress-induced tetragonal martensite, data from the Inorganic Crystal Structure Database (ICSD) were utilized. The cubic austenite phase ($\text{Ni}_2\text{FeGa}$, space group $Fm\bar{3}m$, No. 225, $a = 5.77\text{ \AA}$) was adopted from ICSD entry 187492 ~\cite{Sahariah2013}, whereas the tetragonal martensite phase ($\text{Ni}_{54}\text{Fe}_{19}\text{Ga}_{27}$, space group $I4/mmm$, No. 139, $a = 3.818(6)\text{ \AA}$, $c = 6.496(16)\text{ \AA}$) was taken from ICSD entry 157383 ~\cite{Brown2007}.

The samples for in-situ tensile X-ray diffraction (XRD) and thermomechanical experiment were prepared with the same protocol where a microwire ends were clamped using stainless steel capillaries. The in-situ tensile XRD measurements under applied strain were performed using a Rigaku SmartLab diffractometer equipped with a sealed Cu-anode X-ray tube ($\lambda = 1.5418~\textrm{\AA}$) and an Eulerian cradle. The primary parallel beam was produced by a parabolic multilayer mirror and subsequently collimated using 2.5$^\circ$ Soller slits. No optics were used in the diffracted beam path, and the diffracted intensity was recorded using a HyPix-3000 two-dimensional detector. Sample-detector distance was set to 150~mm. The specimen was mounted in a~custom 3D-printed holder with movable parts that enabled mechanical straining of the sample. The diffraction data were acquired using the 2D scanning mode of the detector. The pole figures were then reconstructed from the recorded frames using a custom-developed algorithm. First, diffraction peaks were automatically detected in the recorded frames. The $2\theta$ angles for resulting pole figures were then determined from the detected peak positions using a~user-defined angular tolerance. Finally, the detected peaks were transformed into angular coordinates, and their contributions to the pole figures were calculated by interpolation.

Thermomechanical experiments on a single microwire were carried out using a Discovery DMA 850 (TA Instruments) in tensile mode. The sample had an initial gauge length of approximately $16.4\text{ mm}$ and a metallic-core diameter of approximately $10.5\ \mu\text{m}$. Stress was calculated from the applied force and the initial metallic cross-sectional area. Thermal cycling under constant stresses of $20$, $50$, and $100\text{ MPa}$ were performed between $-150$ and $150\,^\circ\mathrm{C}$ at a rate of $20\,^\circ\mathrm{C}\cdot\text{min}^{-1}$. The martensite-start temperature was determined from the strain--temperature curve using the tangent method. The transformation strain ($\varepsilon^{\text{tr}}$) was obtained from the strain interval between the martensite-start and martensite-finish temperatures.
The preparation of both austenitic and martensitic regions within a single microwire minimize uncertainties associated with sample-to-sample variation and enables direct phase comparison under identical characterization conditions. This experimental configuration is particularly advantageous for subsequent correlative microscopy studies, where local differences in crystal structure, defect arrangements, and phase-dependent contrast can be directly related to the transformation state.

\section{Results and Discussion}

\begin{figure}[h]
	\centering
	\begin{subfigure}[b]{0.40\linewidth}
		\centering
		\includegraphics[width=\linewidth]{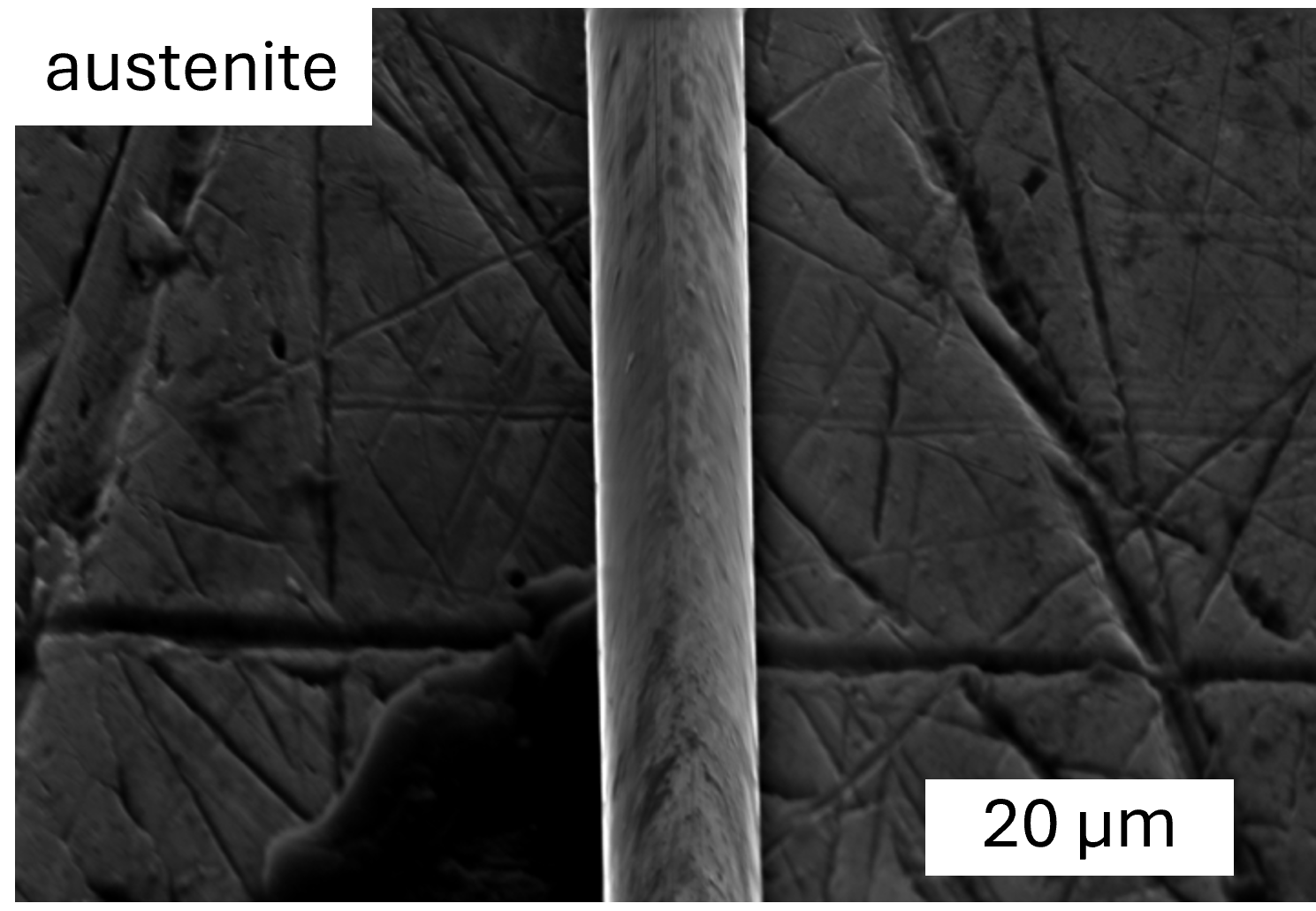}
		\caption{}
		\label{fig:picRD1_a}
	\end{subfigure}
	\begin{subfigure}[b]{0.40\linewidth}
		\centering
		\includegraphics[width=\linewidth]{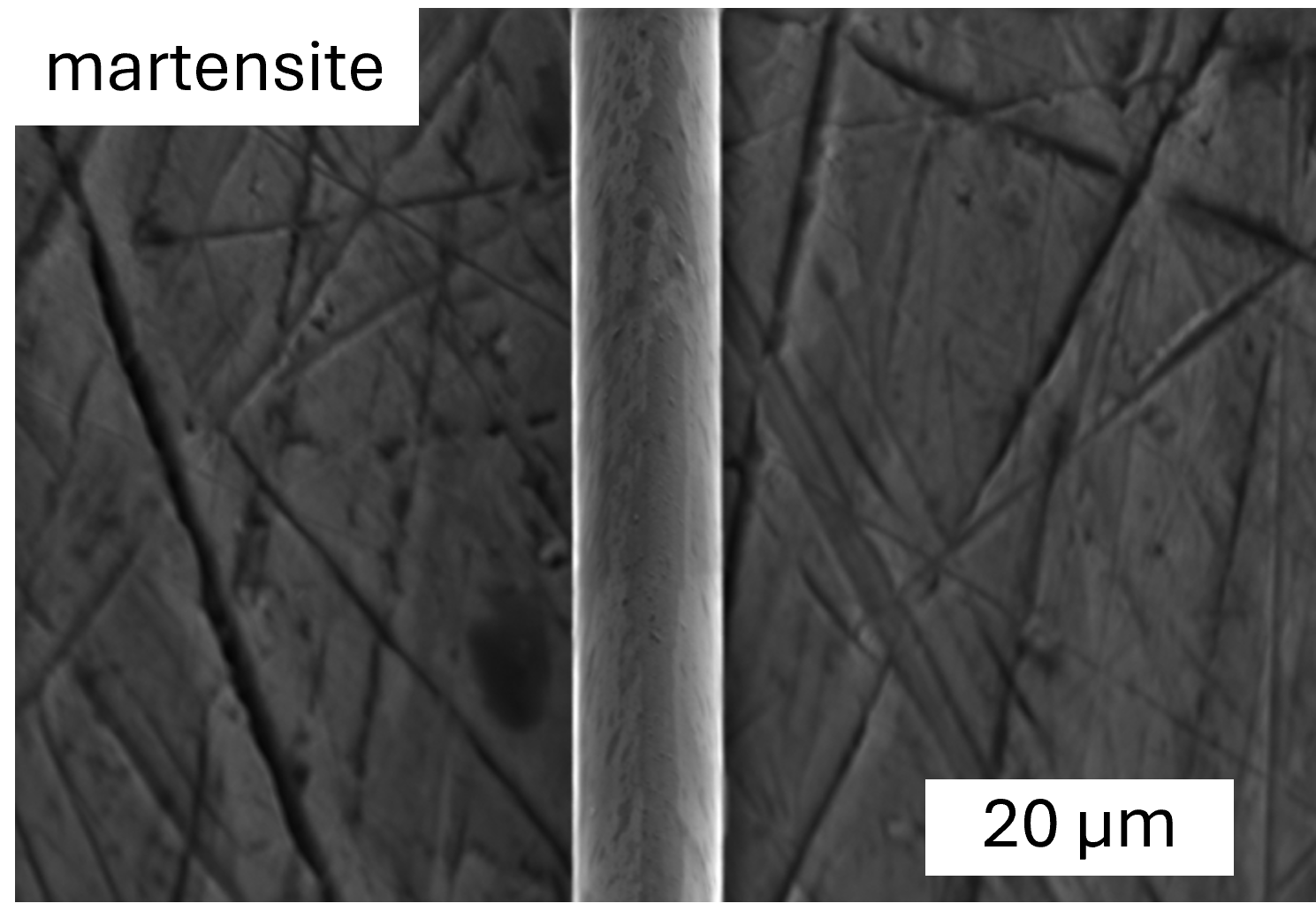}
		\caption{}
		\label{fig:picRD1_b}
	\end{subfigure}
    
	\vspace{1em}
    
	\begin{subfigure}[b]{0.80\linewidth}
		\centering
		\includegraphics[width=\linewidth]{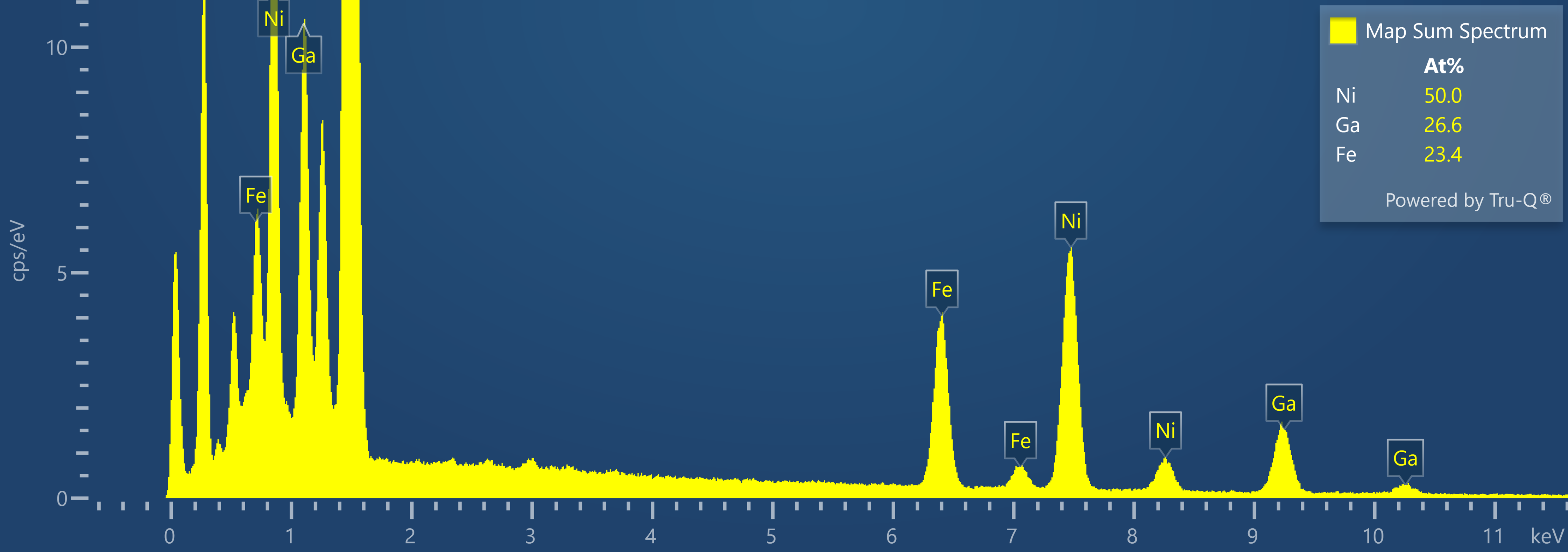}
		\caption{}
		\label{fig:picRD1_c}
	\end{subfigure}
	\caption{SEM micrographs: (a) austenite and (b) martensite regions; and (c) EDX spectrum from the austenite region showing the peaks identification for Ni, Fe, and Ga}
	\label{fig:picRD1}
\end{figure}

Microwire microstructure and chemical composition can be obseved using SEM simultaneously with secondary electron, EDX, and EBSD detectors. Fig.~\ref{fig:picRD1_a} and Fig.~\ref{fig:picRD1_b} shows SEM images of austenitic and strain-induced martensitic regions from a~single sample. The diameter of microwire from austenitic region was measured to be 10.5~$\mu$m, whereas the microwire diameter of strain-induced martensitic region exhibited a~reduced diameter of 10.2~$\mu$m. The observed reduction in diameter of approximately 3~\% reflects the geometric response of the microwire during stress-induced martensitic transformation. EDX performed on the austenitic region (Fig.~\ref{fig:picRD1_c}) revealed a~composition of Ni$_{50.0}$Fe$_{23.4}$Ga$_{26.6}$, which agrees with the nominal alloy composition of Ni$_{50}$Fe$_{23}$Ga$_{27}$ within the typical uncertainty of the EDX method.

\begin{figure}[h]
	\centering
	\begin{subfigure}[b]{0.40\linewidth}
		\centering
		\includegraphics[width=\linewidth]{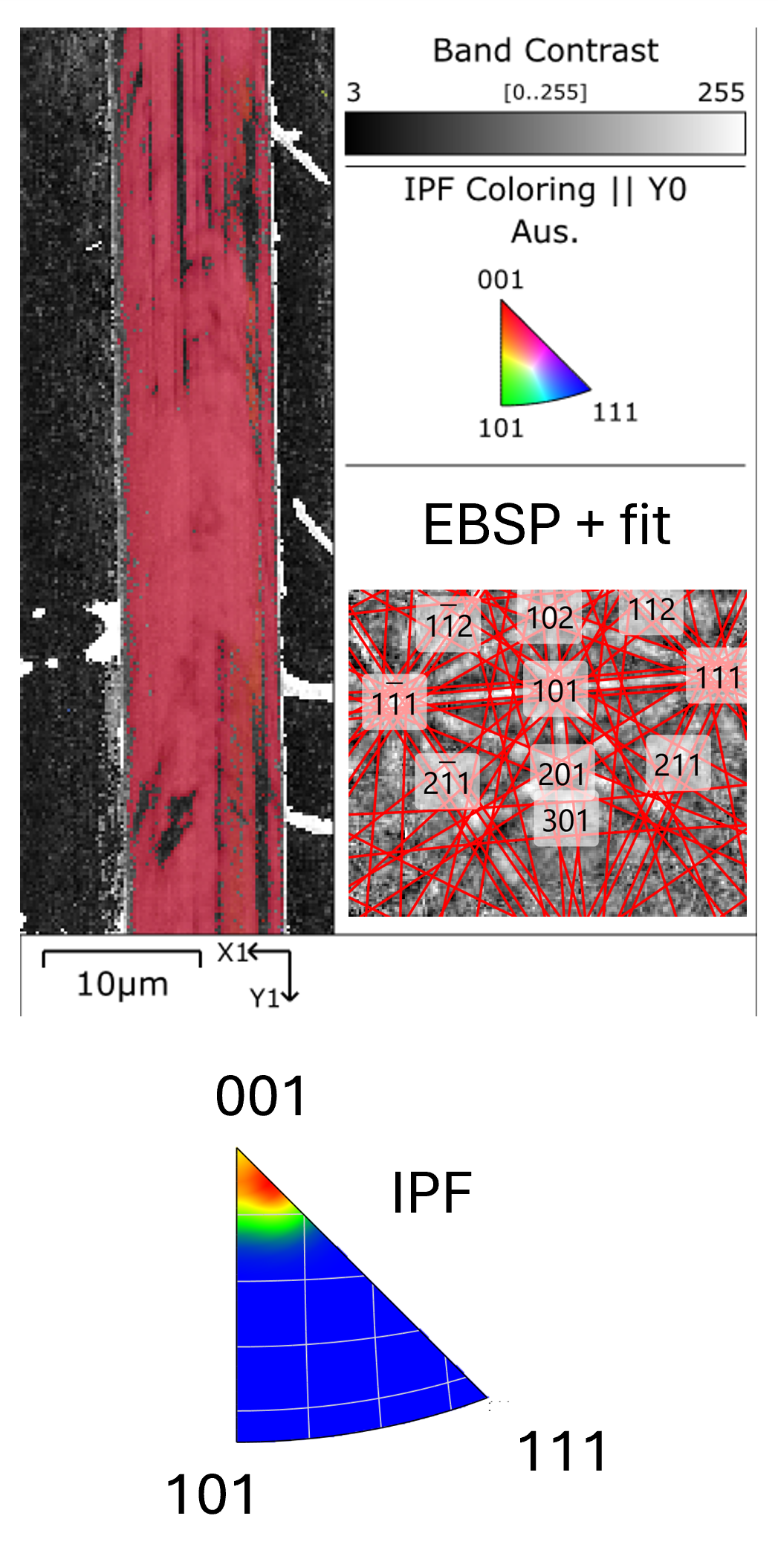}
		\caption{}
		\label{fig:picRD2_a}
	\end{subfigure}
	\begin{subfigure}[b]{0.40\linewidth}
		\centering
		\includegraphics[width=\linewidth]{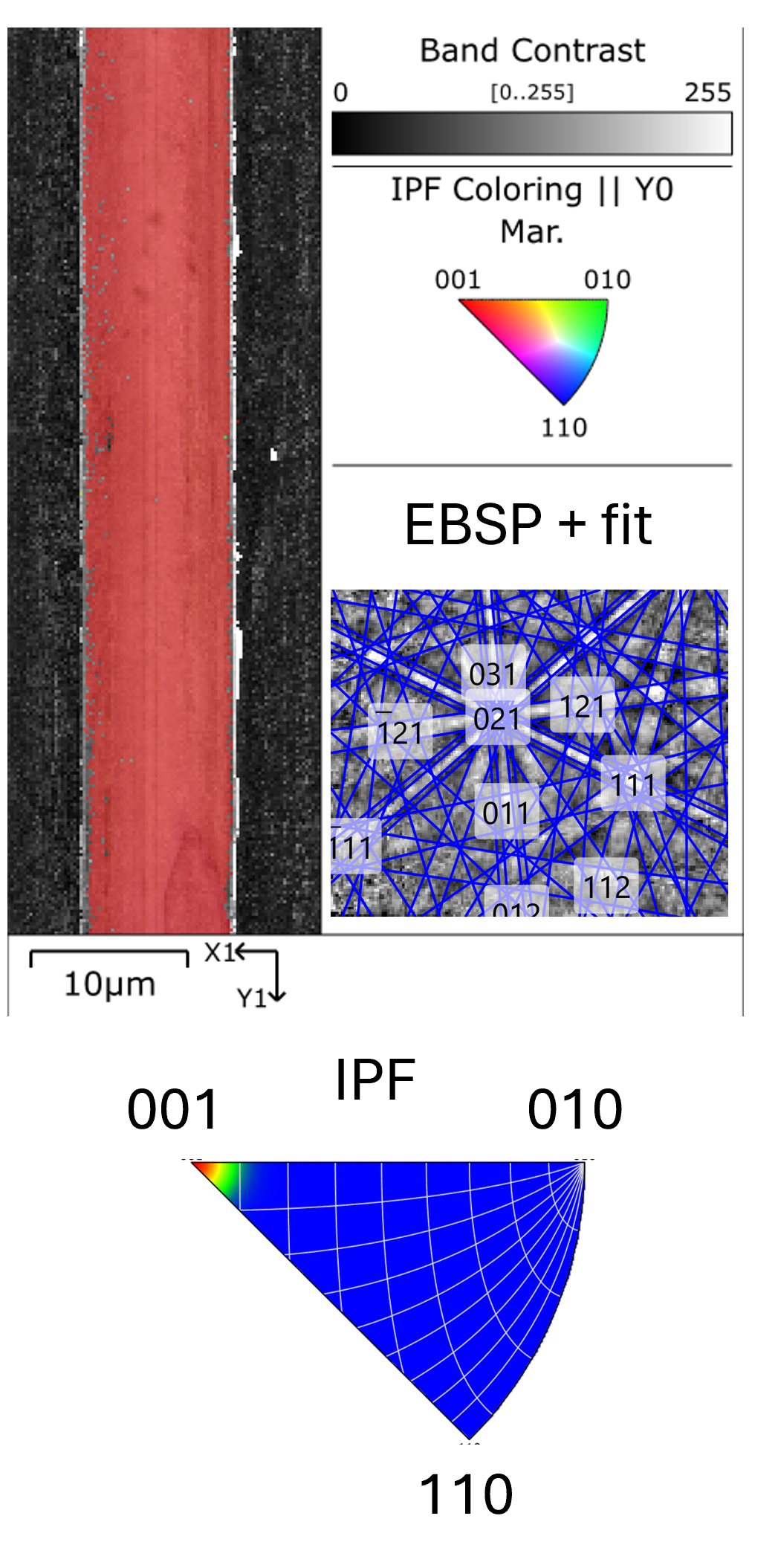}
		\caption{}
		\label{fig:picRD2_b}
	\end{subfigure}
	\caption{Phase orientation mapping along the wire axis ($Y_1$ direction), a representative of EBSD pattern fitting, and corresponding inverse pole figures for (a) austenite and (b) martensite regions}
	\label{fig:picRD2}
\end{figure}

EBSD phase mapping, shown in Fig.~\ref{fig:picRD2}, reveals a clear distinction between the two regions of the microwire. The undeformed region was indexed as cubic austenite (space group $Fm\bar{3}m$, no. 225) with an orientation close to [001] with respect to the wire axis, whereas the strained region was identified as tetragonal martensite (space group $I4/mmm$, no. 139), also exhibiting an orientation near [001] with respect to the wire axis. The observation that both the austenite and the stress-induced martensite exhibit preferred orientations close to [001] suggests the existence of a~favorable crystallographic correspondence during transformation. Such orientation facilitates reversible phase transformation which will be observed during thermomechanical testing.

\begin{figure}[h]
	\centering
	\begin{subfigure}[b]{0.40\linewidth}
		\centering
		\includegraphics[width=\linewidth]{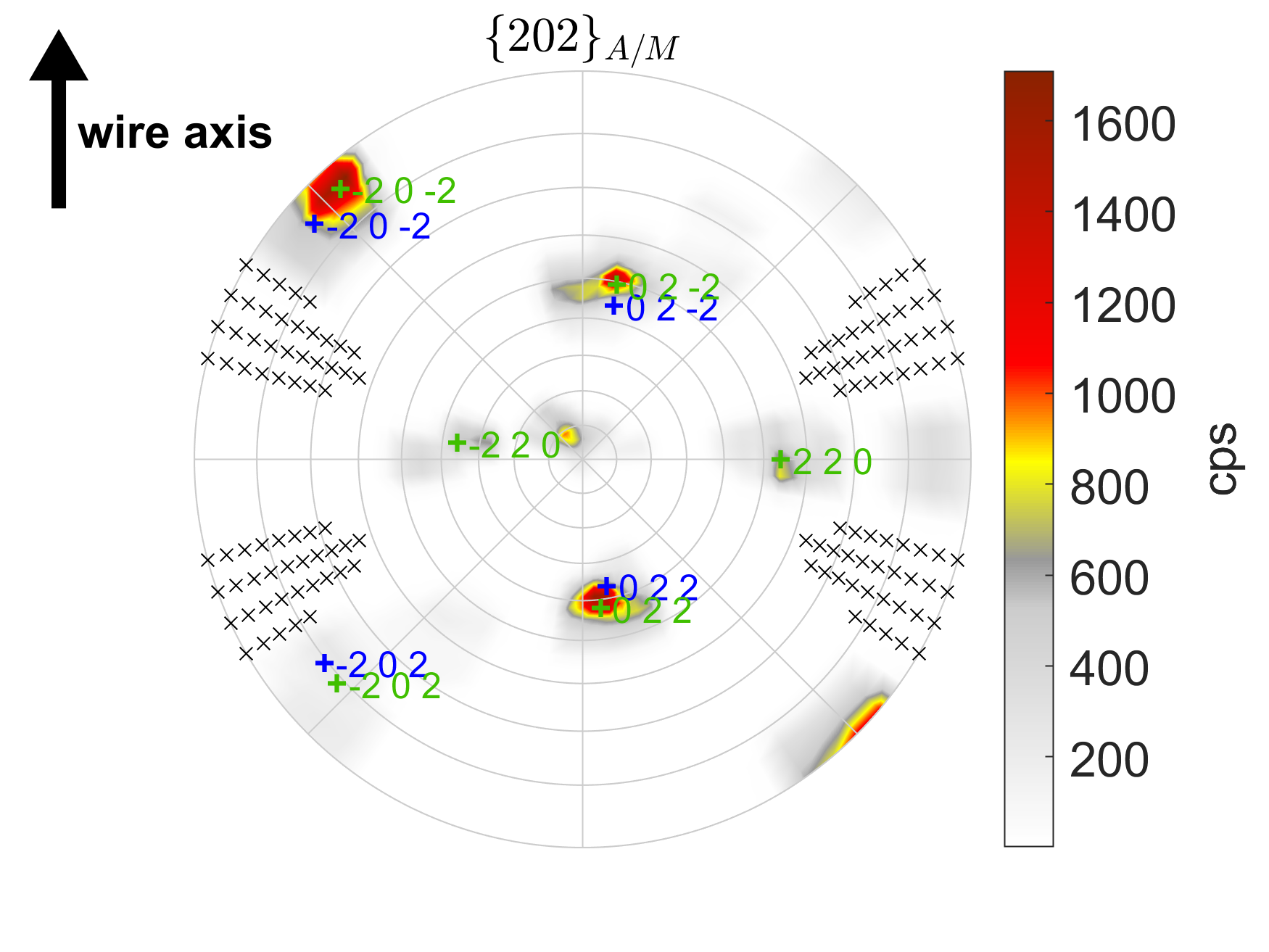}
		\caption{}
		\label{fig:pf_202}
	\end{subfigure}
	\begin{subfigure}[b]{0.40\linewidth}
		\centering
		\includegraphics[width=\linewidth]{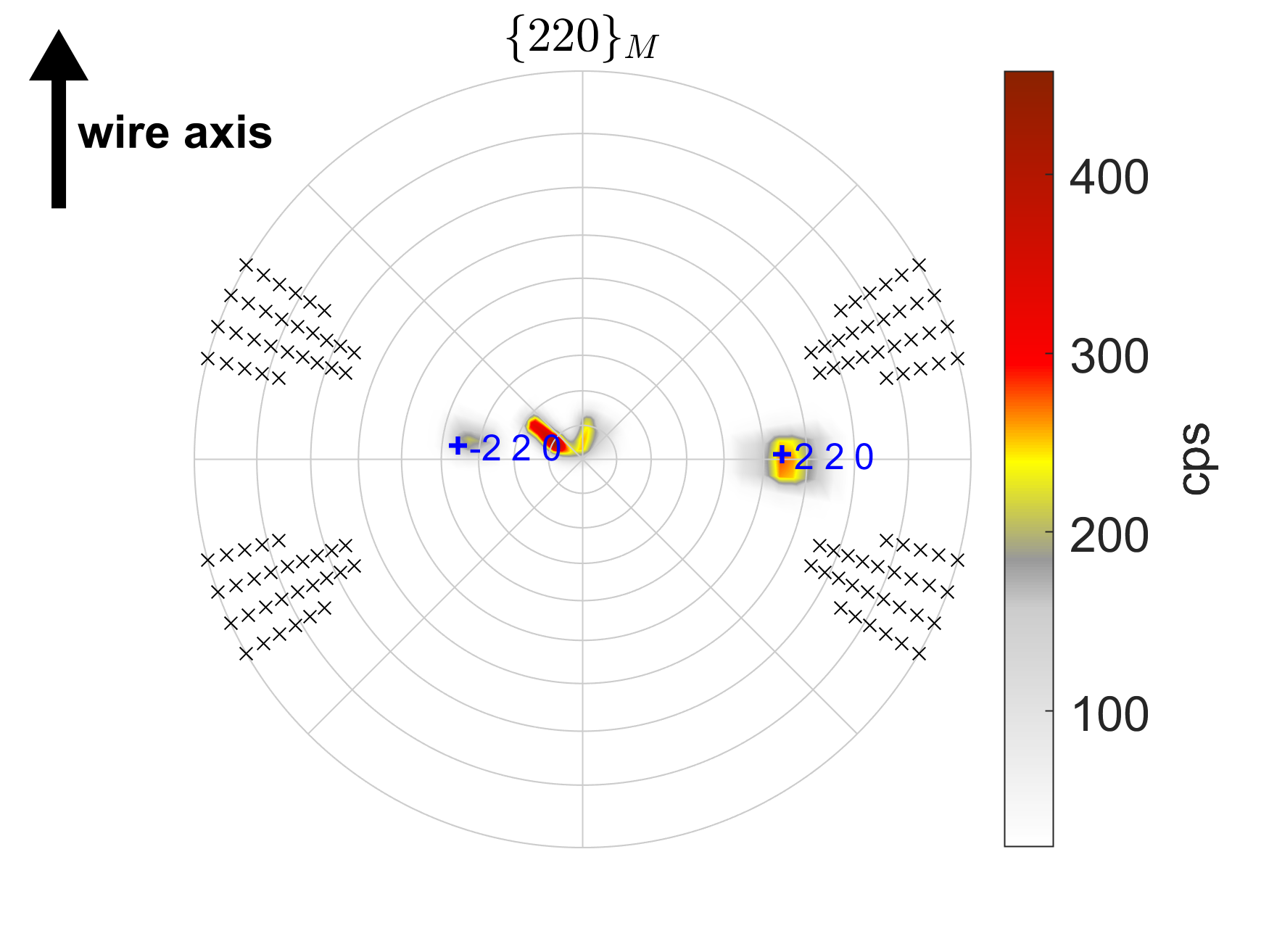}
		\caption{}
		\label{fig:pf_220}
	\end{subfigure}
    
    
	\begin{subfigure}[b]{0.40\linewidth}
		\centering
		\includegraphics[width=\linewidth]{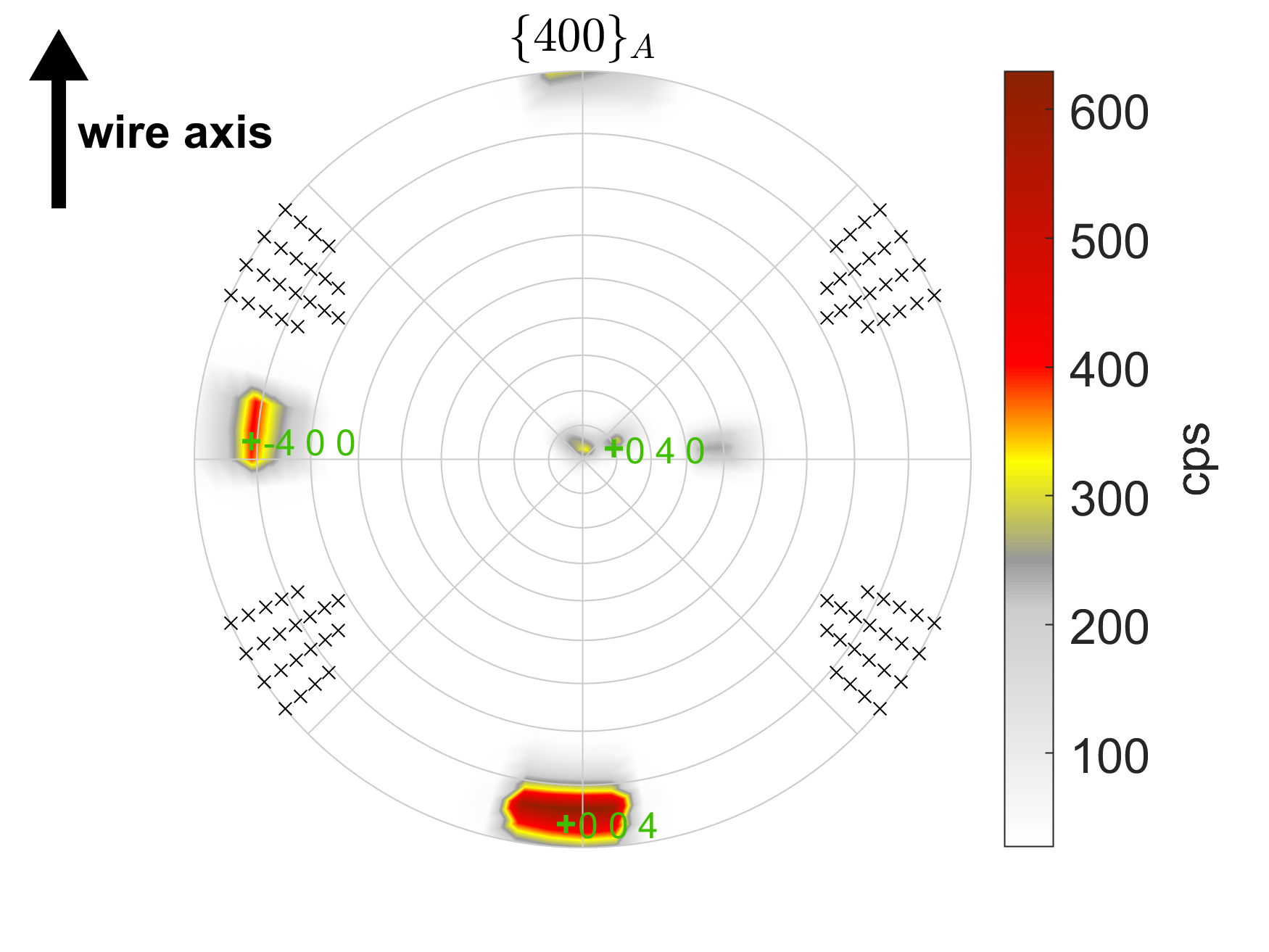}
		\caption{}
		\label{fig:pf_400}
	\end{subfigure}
	\caption{Reconstructed pole figures for the microwire in the initial state: (a) $\{202\}_{A/M}$, (b) $\{220\}_{M}$, and (c) $\{400\}_{A}$. \textit{green crosses} -- simulated positions of the Bragg reflections for the dominant austenitic grain, \textit{blue crosses} -- for the dominant martensitic grain, \textit{black crosses} -- estimated angular regions shadowed by the sample holder during the measurement.}
	\label{fig:picXRDinitial}
\end{figure}

\begin{figure}[h]
	\centering
	\begin{subfigure}[b]{0.40\linewidth}
		\centering
		\includegraphics[width=\linewidth]{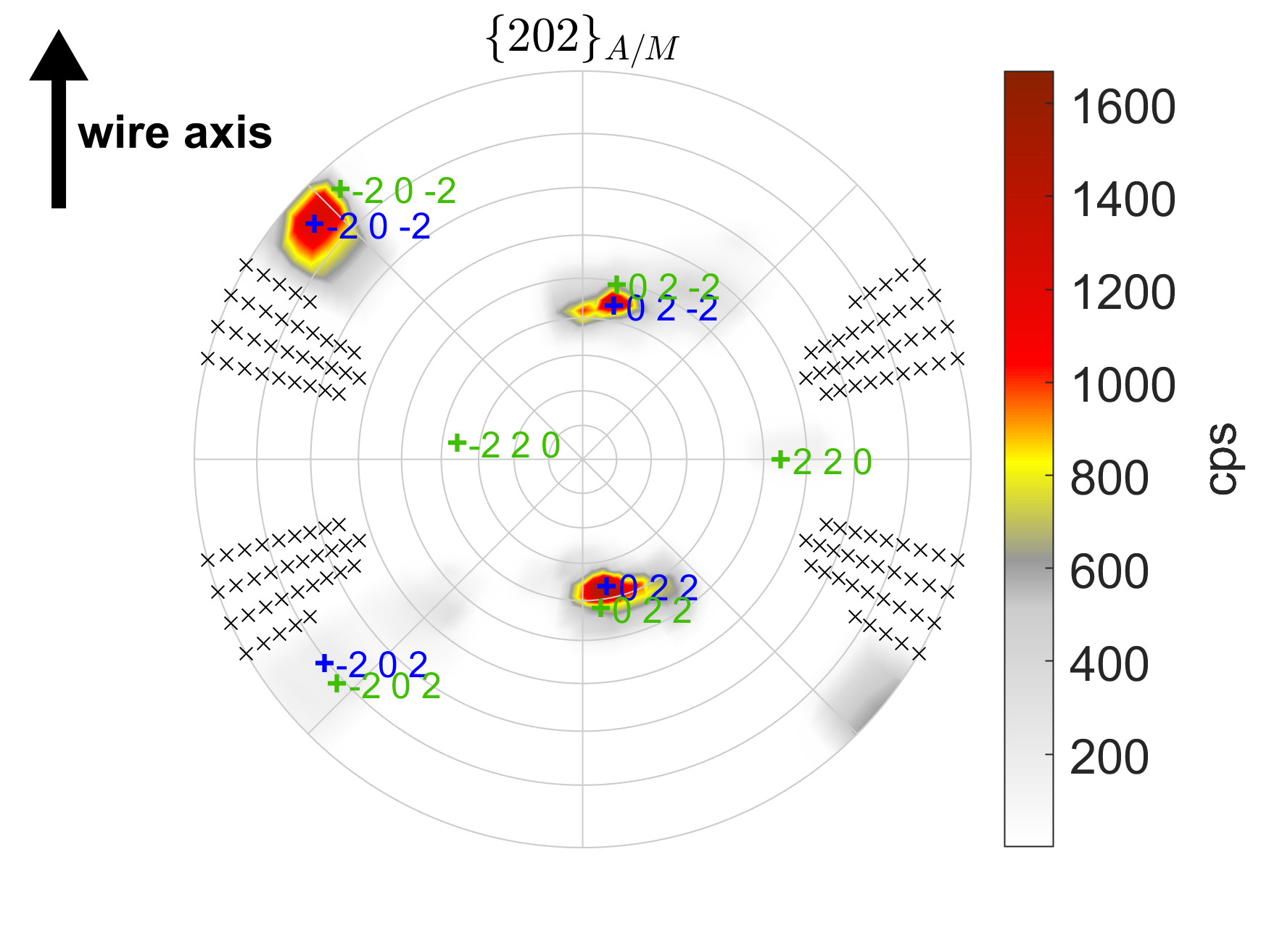}
		\caption{}
		\label{fig:pf_202_strained}
	\end{subfigure}
	\begin{subfigure}[b]{0.40\linewidth}
		\centering
		\includegraphics[width=\linewidth]{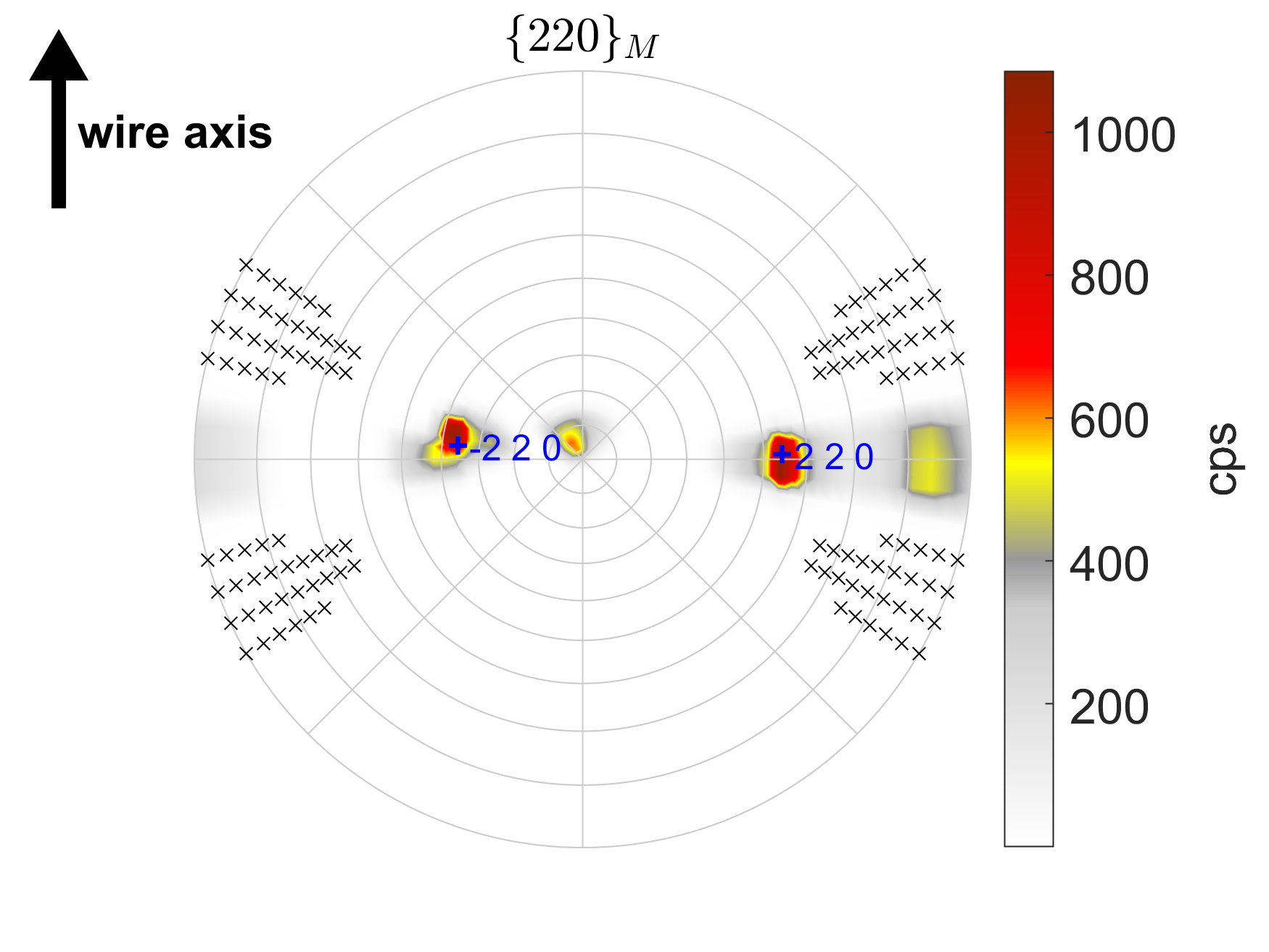}
		\caption{}
		\label{fig:pf_220_strained}
	\end{subfigure}
    
    
	\begin{subfigure}[b]{0.40\linewidth}
		\centering
		\includegraphics[width=\linewidth]{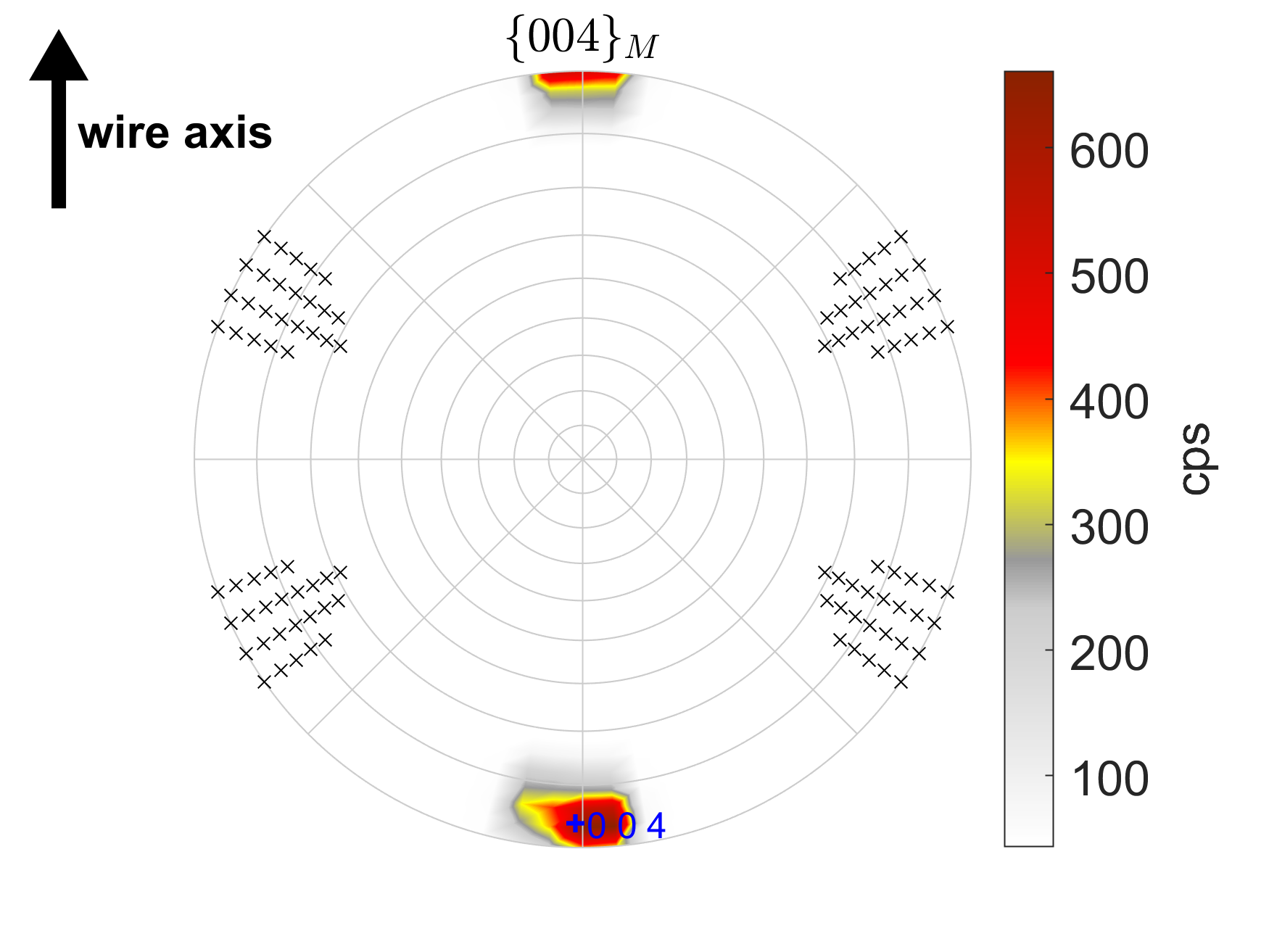}
		\caption{}
		\label{fig:pf_004_strained}
	\end{subfigure}
	\begin{subfigure}[b]{0.40\linewidth}
		\centering
		\includegraphics[width=\linewidth]{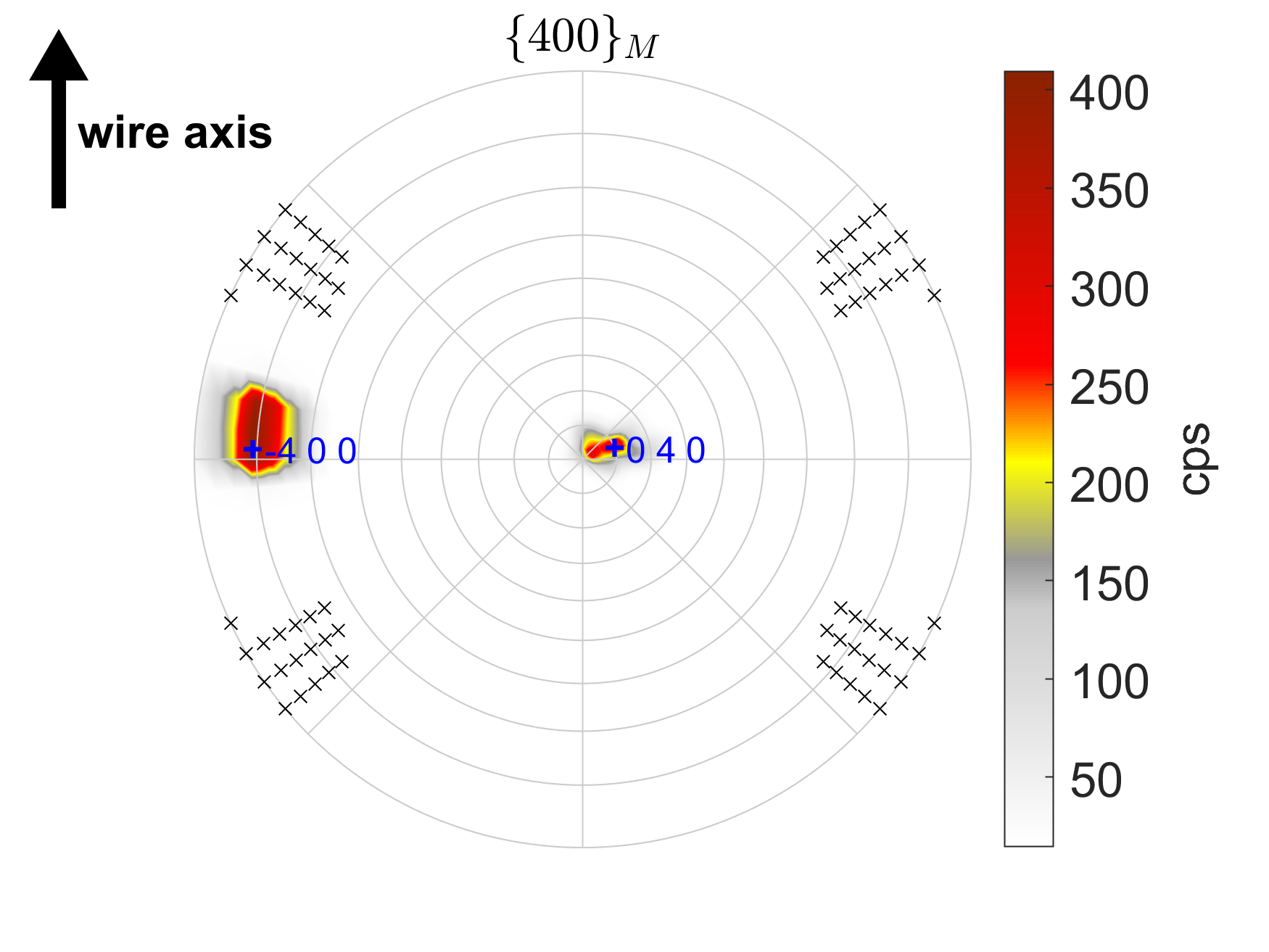}
		\caption{}
		\label{fig:pf_400_strained}
	\end{subfigure}
	\caption{Reconstructed pole figures for the microwire in the strained (around $11\,\%$) state: (a) $\{202\}_{A/M}$, (b) $\{220\}_{M}$, (c) $\{004\}_{M}$, and (d) $\{400\}_{M}$. \textit{Green crosses} denote simulated positions of the Bragg reflections for the dominant austenitic grain, \textit{blue crosses} for the dominant martensitic grain, and \textit{black crosses} the estimated angular regions shadowed by the sample holder during the measurement.}
	\label{fig:picXRDstrained}
\end{figure}

Characterization from EBSD shows that the transformation from cubic austenite to tetragonal martensite in ferromagnetic Ni-Fe-Ga alloys is accompanied by a lattice distortion simultaneously lowers the crystal symmetry while preserving the overall crystallographic correspondence. This martensitic transformation was also confirmed by in-situ tensile XRD measurements. The reconstructed pole figures in Fig.~\ref{fig:picXRDinitial} and Fig.~\ref{fig:picXRDstrained} show the change in Bragg reflections corresponding to the austenitic cubic structure (space group $Fm\bar{3}m$, no. 225) into reflections corresponding to the tetragonal martensite. For the crystallographic description of the martensite, the austenitic axes were inherited. This description is also more suitable for the calculation of the theoretical transformation strain. The martensitic unit cell in this description is bigger and rotated by 45$^\circ$ about the $z$-axis relative to the unit cell used for EBSD indexing. Consequently, to obtain the corresponding lattice parameters, the lattice parameter $a$ used for EBSD indexing must be multiplied by $\sqrt{2}$. Ideally, the Bragg reflections within each pole figure would exhibit similar intensities. However, the observed intensity variations primarily reflect changes in the irradiated sample area during the pole figure measurement.

Due to experimental limitations, the in-situ tensile XRD does not allow measurements to be performed at exactly $0\,\%$ strain (fully austenitic state), as a tensile preload was required to maintain a microwire straight and prevent sample movement during pole figure measurement. Consequently, a small fraction of martensite was still present in the low-strain condition (see Fig.~\ref{fig:picXRDinitial} and pole figure $\{220\}_M$). Despite this phase coexistence, the diffraction peaks were sufficiently resolved to enable reliable determination of the lattice parameters of the individual phases. The measurement in the strained state was performed at a tensile strain of approximately $11.0\,\%$. Strain values approaching the maximum strain of $15.0\,\%$ pose a risk of breaking the wire.

The austenitic phase was indexed as a cubic structure with a lattice parameter of $a_A = 5.72(4)~\text{ \AA}$. The strain-induced martensitic phase exhibited a~tetragonal structure with the lattice parameters $a_M=~5.38(1)~\text{ \AA}$ and $c_M=~6.60(6)~\text{ \AA}$ using coordinates inherited from austenite phase. This way we can directly calculate the tetragonal distortion of martensite phase upon martensitic transformation
\[\frac{c_M}{a_M}=1.23.\]
This pronounced tetragonality is consistent with large transformation strain observed which is given by difference between lattice constants of austenite and martensite in experiment. In-situ tensile XRD measurement also confirms the preferred [001] orientation of the wire since the Bragg reflections 004 in both phases lie approximately along the wire direction (the vertical axis in the pole figures). Therefore, the lattice-derived transformation strain is

$\varepsilon^{\text{tr}} = \left( \frac{c_M}{a_A} - 1 \right) \approx 15.4\,\%$.

\begin{figure}[h]
	\centering
	\includegraphics[width=0.60\linewidth]{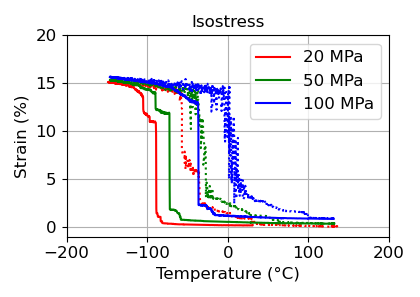}
	\caption{Thermal cycles under constant 20, 50, and 100~MPa tensile stress (solid lines = cooling and dotted lines = heating)}
	\label{fig:picRD5}
\end{figure}

Moreover, the thermal cycling under constant stress in Fig.~\ref{fig:picRD5} reveals an increase in transformation temperatures with increasing applied stress. It can also be seen that the length of martensite is less depended on the applied stress than the length of austenite. To obtain more quantitative results, martensite temperature and experimental transformation strain are calculated (\ref{Appendix}).

Table~\ref{tab:stress_ms_strain} shows the applied stress versus martensite start temperature. Using the martensite start temperature, the stress-temperature slope is approximately 1.52~MPa$/^{\circ}$C, close to the value of 1.42~MPa$/^{\circ}$C reported for [001]-oriented Ni$_{54}$Fe$_{19}$Ga$_{27}$ single crystals~\cite{Pataky2015}. This observed stress sensitivity confirms the relatively small Clausius--Clapeyron slope characteristic of Ni--Fe--Ga alloys. The agreement suggests that the microwires exhibit a transformation behavior similar to highly ordered bulk materials.

\begin{table}[htbp]
\centering
\caption{Experimental martensite start temperature ($M_s$) and transformation strain as a function of applied stress.}
\label{tab:stress_ms_strain}
\begin{tabular}{ccc}
\toprule
Stress (MPa) & $M_s$ ($^\circ$C) & Transformation strain (\%) \\
\midrule
20  & -90 & 14.0 \\
50  & -73 & 13.3 \\
100 & -37 & 13.2 \\
\bottomrule
\end{tabular}
\end{table}

The influence of applied stress on the experimental transformation strain can be seen in Table~\ref{tab:stress_ms_strain}. Considering the effect of applied stress on the length of martensite and austenite, the trend of increasing applied stress from $20\text{ MPa}$ to $100\text{ MPa}$ leads to decreasing of transformation strain from $14.0\,\%$ to $13.2\,\%$. Moreover, these transformation strain of $13.2--14.0\,\%$, remaining in close agreement with the theoretical lattice-derived strain ($15.4\,\%$) where is the obtained from the in-situ tensile XRD measurement.

Since Ni${50}$Fe${23}$Ga$_{27}$ is close to the stoichiometric composition of the ordered intermetallic Ni$_2$FeGa phase~\cite{Nath2015}, the EBSD and in-situ tensile XRD results confirm that the microwire investigated in this study does not contain the $\gamma$-phase. In contrast to typical bulk Ni–Fe–Ga alloys, the microwire exhibits an experimental transformation strain of up to $14.0\,\%$, possibly owing to its small diameter of about 10~$\mu$m. Despite the small diameter of the microwire, the required mN-scale force is sufficiently low to allow the strained sample to be secured with silver paste for EBSD characterization. Nevertheless, the microwire remains highly susceptible to fracture during sample handling.

The combination of a large experimental transformation strain ($14.0\,\%$) and a relatively small Clausius--Clapeyron slope (1.52~MPa$/^{\circ}$C) is particularly advantageous for actuator applications. A low Clausius--Clapeyron slope indicates that comparatively small stresses are required to modify phase stability, while the large crystallographic transformation strain provides a substantial mechanical response. Consequently, the investigated microwires can generate large recoverable deformations under relatively moderate loading conditions.

\section{Conclusion}
The crystallographic evolution and thermomechanical behavior of a Taylor--Ulitovsky Ni${50}$Fe${23}$Ga$_{27}$ microwire were investigated using EBSD, in-situ tensile XRD, and thermomechanical characterization. The austenite was identified as a cubic phase (space group $Fm\bar{3}m$, No.~225) with a lattice parameter of $a_A = 5.72(4)$~\AA{} and a preferred [001] orientation along the wire axis. Under tensile loading, the strain-induced martensite with a tetragonal structure ($a_M = 5.38(1)$~\AA{} and $c_M = 6.60(6)$~\AA{}) formed while retaining a [001] orientation parallel to the wire axis. The corresponding crystallographically calculated transformation strain of approximately $15\,\%$ is in good agreement with the experimentally measured value of about $14\,\%$, confirming the martensitic transformation mechanism. Furthermore, thermomechanical analysis estimates a Clausius-Clapeyron slope of approximately 1.52~MPa$/^{\circ}$C, which is comparable to values reported for [001]-oriented Ni-Fe-Ga single crystals. These results demonstrate that the investigated microwire exhibits a highly oriented martensitic transformation associated with a large recoverable transformation strain.

\section*{CRediT authorship contribution statement}
\textbf{Limpat Nulandaya:} Investigation, Writing - Original Draft
\textbf{Petr Cejpek:} Investigation, Writing - Review \& Editing
\textbf{Ladislav Straka:} Supervision, Writing - Review \& Editing, Validation
\textbf{Oleg Heczko:} Supervision, Writing - Review \& Editing, Funding Acquisition
\textbf{Rastislav Varga:} Supervision, Writing - Review \& Editing, Resources, Funding Acquisition

\section*{Declaration of generative AI and AI-assisted technologies in the manuscript preparation process}
During the preparation of this work, the authors used Microsoft Copilot (GPT 5.6 Sol) and Claude (Anthropic) phrasing improvements in a few instances. After using these tools, the authors reviewed and edited the content as needed and take full responsibility for the final content of the published article.

\section*{Declaration of Competing Interest}
The authors declare that they have no known competing financial interests or personal relationships that could have appeared to influence the work reported in this paper.

\section*{Acknowledgements}
This work was financially supported by the Operational Programme Johannes Amos Comenius (OP JAK) of the Ministry of Education, Youth and Sport of the Czech Republic, within the frame of project Ferroic Multifunctionalities (FerrMion) [project No. CZ.02.01.01/00/22\_008/0004591], co-funded by the European Union. This work was also co-funded by the European Union through the the Technology and Applications for Competitiveness Operational Programme (OP TAK) project “SMART actuators based on micro wires from Heusler alloys” (SMARTACT), project No. CZ.01.01.01/01/24\_063/0006741. Researcher Petr Cejpek conducts his research under the Marie Skłodowska-Curie Actions -- COFUND project, which is co-funded by the European Union (Physics for Future -- Grant Agreement No. 101081515).

\section*{Appendix: Determination of martensite start temperature ($M_s$), transformation strain, and stress-temperature slope}\label{Appendix}

\begin{figure}[h]
	\centering
	\begin{subfigure}[b]{0.40\linewidth}
		\centering
		\includegraphics[width=\linewidth]{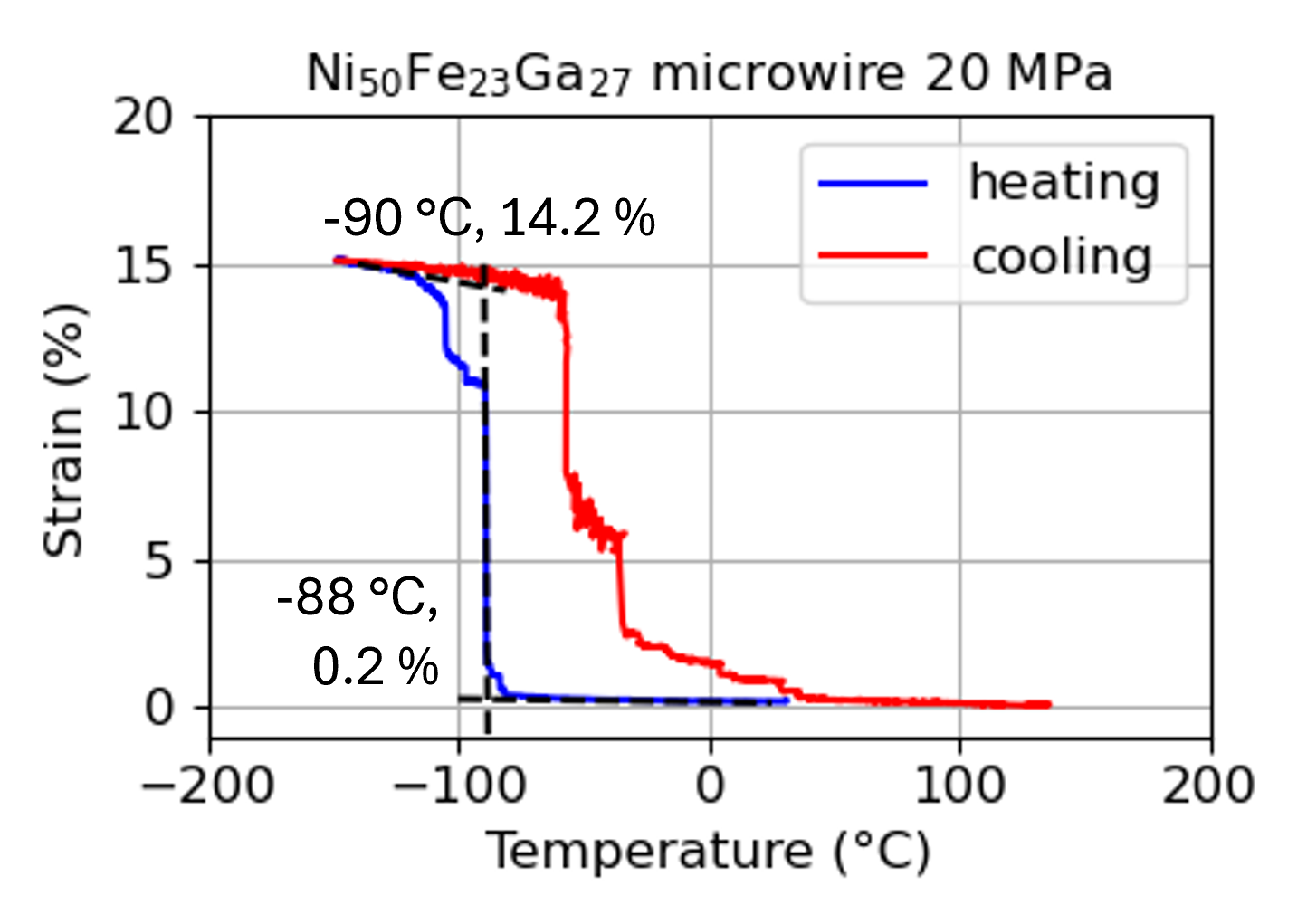}
		\caption{}
		\label{fig:A7a}
	\end{subfigure}
	\begin{subfigure}[b]{0.40\linewidth}
		\centering
		\includegraphics[width=\linewidth]{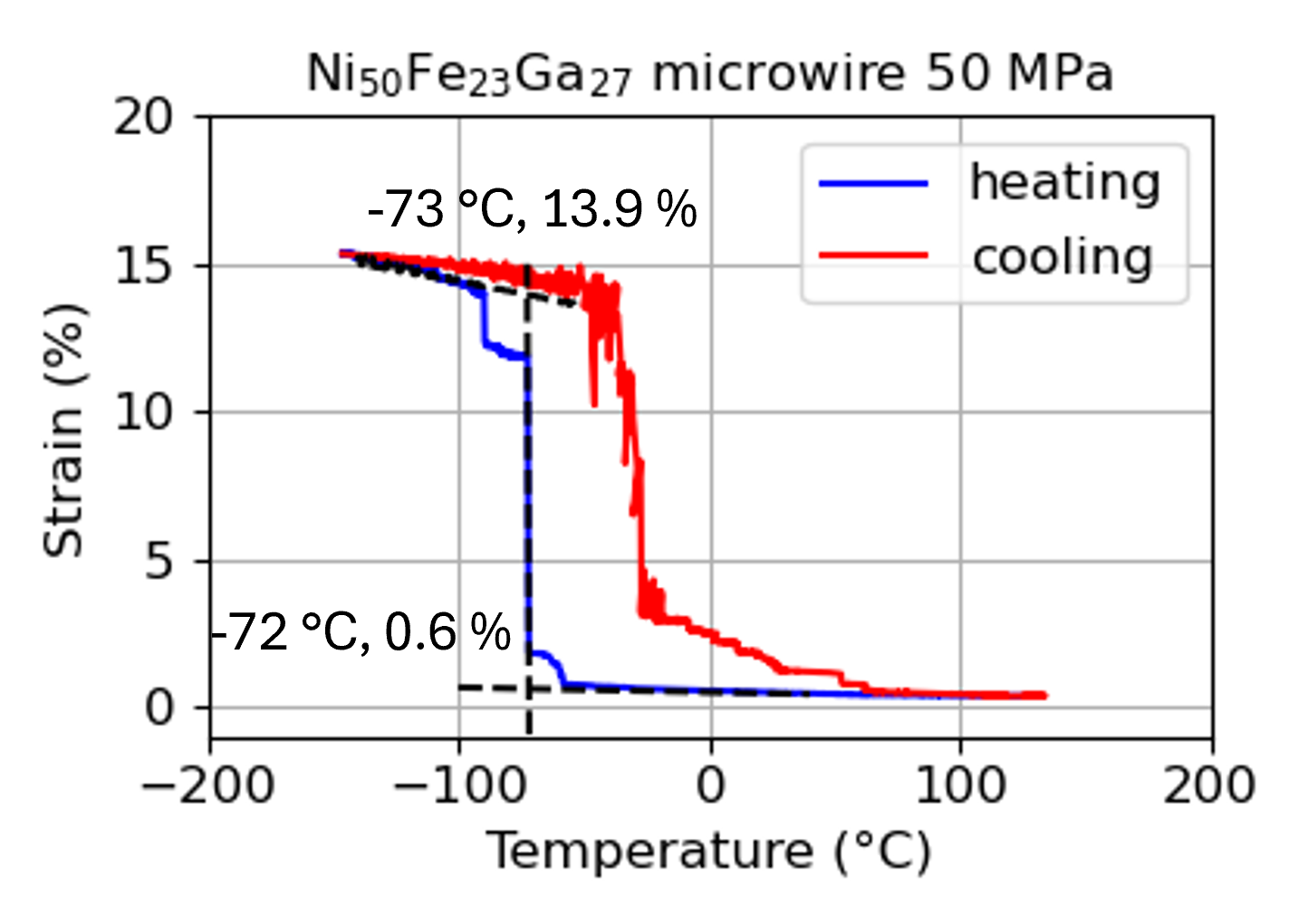}
		\caption{}
		\label{fig:A7b}
	\end{subfigure}
    
    
	\begin{subfigure}[b]{0.40\linewidth}
		\centering
		\includegraphics[width=\linewidth]{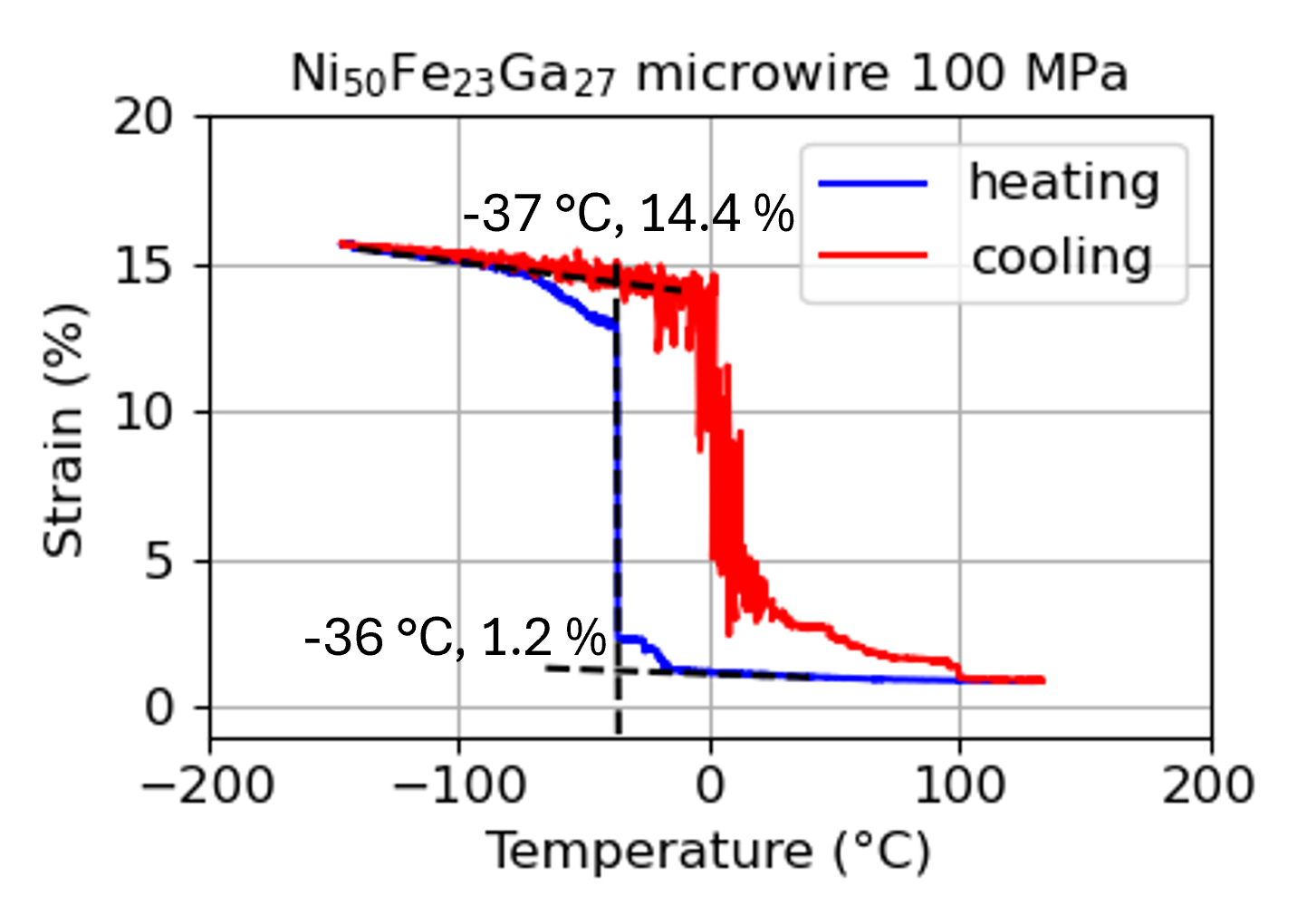}
		\caption{}
		\label{fig:A7c}
	\end{subfigure}
	\begin{subfigure}[b]{0.40\linewidth}
		\centering
		\includegraphics[width=\linewidth]{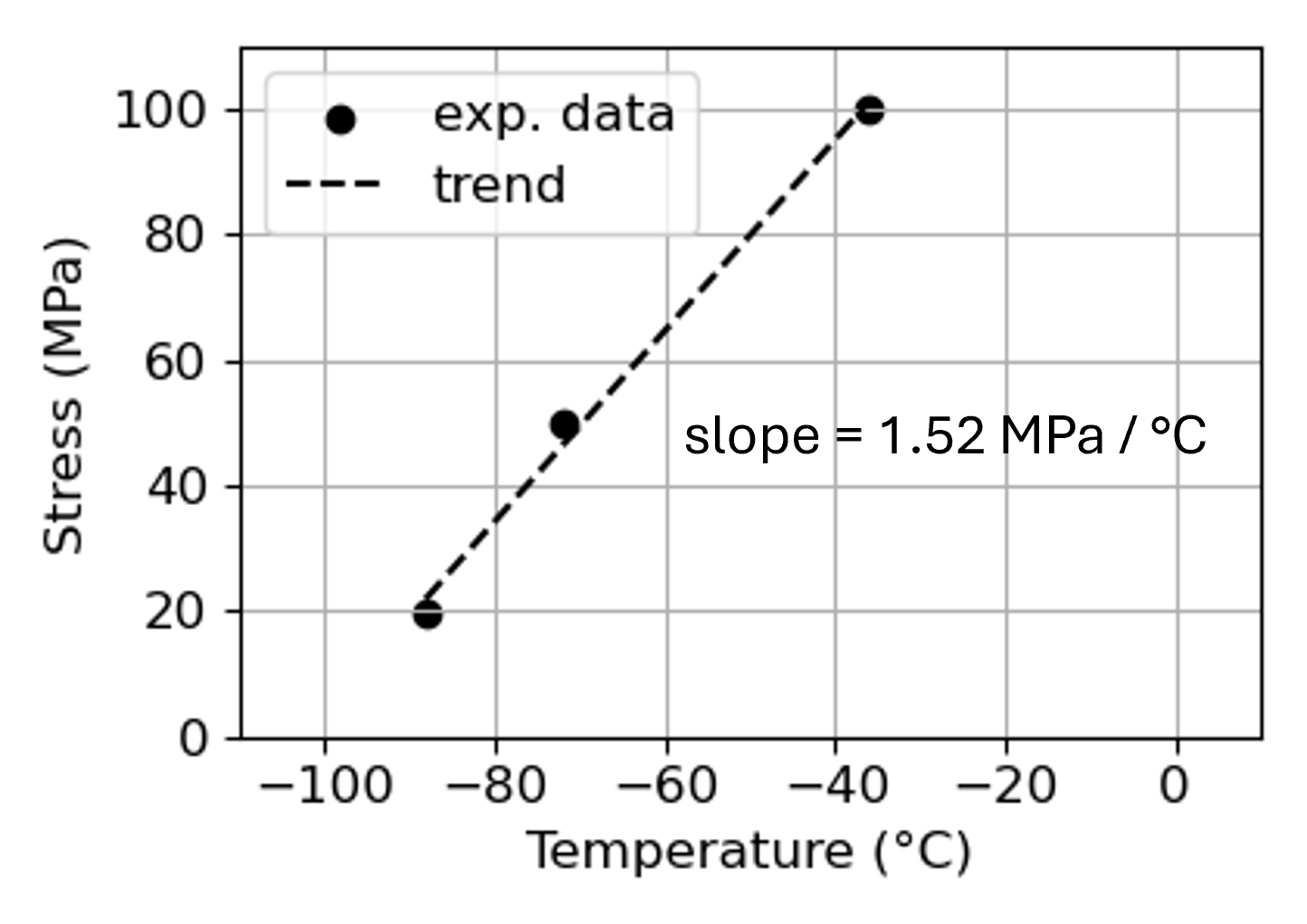}
		\caption{}
		\label{fig:A7d}
	\end{subfigure}
	\caption{Determination of martensite start temperature ($M_s$) and transformation strain from thermal cycle under: (a) 20 MPa, (b) 50 MPa, and (c) 100 MPa; and (d) the determination of stress-temperature slope}
	\label{fig:appendix}
\end{figure}

The determination of martensite start temperature ($M_s$) and transformation strain from thermal cycle under 20~MPa (Fig.~\ref{fig:A7a}), 50~MPa (Fig.~\ref{fig:A7b}), and 100~MPa (Fig.~\ref{fig:A7c}) are following the tangen method. The Clausius-Clapeyron slope is determined from the transformation stress vs. martensite start temperature ($M_s$) curve (\ref{fig:A7d}).

\bibliographystyle{model1-num-names}

\bibliography{cas-refs}



\end{document}